\documentclass[a4paper,12pt]{article}
\usepackage{float}
\usepackage{tabularx}
\usepackage{amsmath}
\usepackage{amsfonts}
\usepackage{adjustbox}
\usepackage{multirow}
\usepackage{color}
\usepackage{longtable}
\usepackage{pdflscape}
\usepackage{rotating}
\usepackage{xcolor}
\usepackage[colorlinks = true,
            linkcolor = blue,
            urlcolor  = blue,
            citecolor = blue,
            anchorcolor = blue]{hyperref}
\usepackage[margin=1in,footskip=0.25in]{geometry}
\title{$n\alpha$ Elastic Scattering: An Application of Variable Phase Approach to Local Potential}
\author{Anil Khachi$^*$\\\\
Chandigarh Group of Colleges Jhanjeri, Mohali, Punjab, India- 140307\\  Chandigarh Engineering College, Department of Applied Sciences}
\begin{document}
\maketitle
\begin{abstract}
\noindent Distance-dependent phase shifts, amplitude functions, and radial wave functions
for neutron-alpha elastic scattering are studied using the Variable Phase
Approach. The microscopic KKNN potential is employed to calculate scattering
properties for the $S_{1/2}$, $P_{3/2}$, and $P_{1/2}$ partial waves over a range
of laboratory energies. The variable phase equations are solved numerically using
a fifth-order Runge--Kutta method, allowing a direct examination of how the
nuclear interaction generates the scattering phase within the finite interaction
region. The results exhibit physically consistent behavior of the phase shifts
and yield well-behaved amplitude and wave functions. This study demonstrates
that the Variable Phase Approach provides a physically transparent and reliable
framework for describing neutron--alpha elastic scattering and for applications
in inverse scattering problems.
\end{abstract}

\section{Introduction}
Under ``standard approach" main objective of theoretical physicst is to calculate the
wavefunction from which other scattering properties like phase shifts, low energy scattering
parameters, static properties etc., can be obtained. It is to be noted that the experimentalists do not get wavefunction as an output of an experiment, rather one gets outputs like differential cross sections and total cross-sections, transition energies etc., which are a result of the interaction processes. Phase shifts are further related to the cross sections. The interaction is portrayed from the experimental outputs in form of V(r) vs r (fm) plots. According to Schwinger, Landau and Smorodinsky \cite{1} ``The meeting place between theory and experiment is not the phase shifts themselves but the value of the variational parameters implied by phase shifts".
\newline There are various techniques that can be used to obtain the scattering phases by solving the
Schrödinger equation like: Born approximation, Brysk's approximation,and other successive approximation techniques. In our earlier papers we have applied PFM or VPA for studying n-p \cite{2,3,4}, n-d \cite{5} and n-$\alpha$ \cite{6}  using Morse potential. Knowing that using PFM, phase
shifts of the nucleon-nucleon scattering can be obtained using different types of potentials (high
precision phenomenological potentials) and interaction models between two nucleons. This paper is an extension of our recent published work \cite{8} where we have considered Morse potential \cite{7} as nuclear interaction for obtaining scattering phase shifts (SPS) for various $\ell$ channels in \textit{np} scattering. The current paper deals with
using the obtained parameters \cite{8} and using those parameters we have obtained various functions like $\delta$ vs r,$A$ vs r and $u$ vs r.
\section{Methodology}
\subsection{Interaction Potential}
\begin{figure}[ht!]
    \centering
    \includegraphics[scale=0.8]{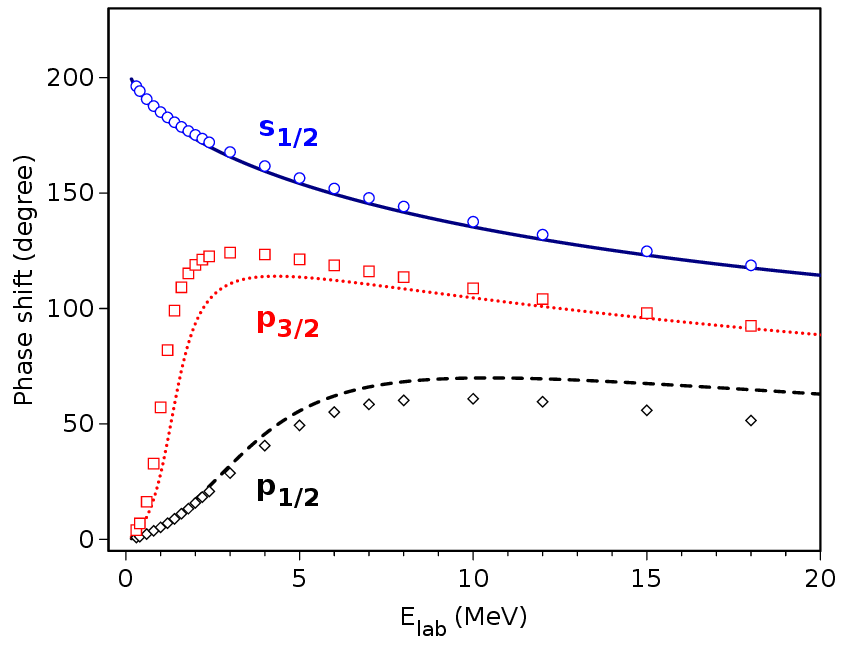}
     \includegraphics[scale=0.85]{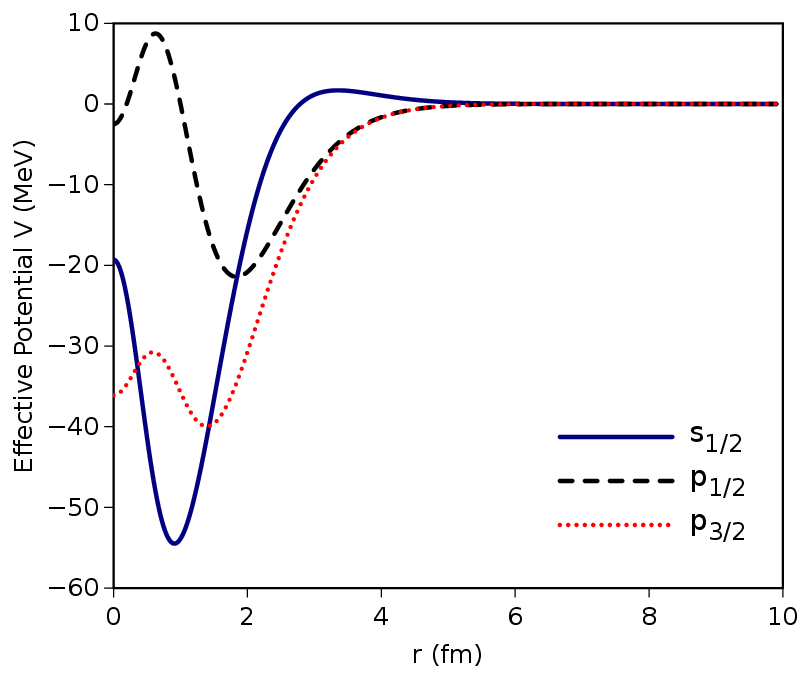}
    \caption{Scattering phase shifts as a function of laboratory energy for the
$S_{1/2}$, $P_{1/2}$, and $P_{3/2}$ partial waves of neutron--alpha elastic
scattering, obtained using the Variable Phase Approach with the KKNN
potential. The corresponding effective interaction potentials for each
channel are also shown. The differences between the partial waves reflect
the roles of the central interaction, centrifugal barrier, and spin-orbit
coupling in the n-$\alpha$ system.}
    \label{sps_pot}
    \end{figure}

We have used the so-called microscopic
KKNN potential \cite{odsu} given by Kanada \textit{et. al}
\begin{align}\nonumber
V_{n-\alpha(r)}&=\sum_{i=1}^2 V_i^C exp(-\mu_i^C r^2)\\& \nonumber+(-1)^{\ell}\sum_{i=1}^3 V_{\ell i}^C exp(-\mu_{\ell i}^C r^2)\\& \nonumber+(\ell.s)\bigg[  V^{\ell s} exp(-\mu^{\ell s}r^2)\\&+\{1+0.3(-1)^{\ell-1}\}\sum_{i=1}^2 V_{\ell i}^{\ell s} exp(-\mu_{\ell i}^{\ell s}r^2)\bigg]
\end{align}
The parameters used in KKNN are given in table \ref{KKNN}. We will now substitute the KKNN potential into the main variable phase equation and the obtained scattering phase shifts for $S_{1/2}, p_{3/2} \& p_{1/2}$ states are shown in figure \ref{sps_pot} respectively. 

\begin{table}[]
\centering
\caption{Parameters of the $n-\alpha$ system using KKNN potential \cite{odsu}.}
\begin{tabular}{lllll}
\hline \hline
           & $V^C$ (MeV)    & $\mu^C$(fm$^{-2}$)   & $V_{\ell}^C$ (MeV)    & $\mu_{\ell}^C$(fm$^{-2}$)    \\ \hline  
Central    & -96.3 & 0.36 & 34    & 0.20  \\ 
           & 77.0  & 0.90 & -85   & 0.53  \\ 
           &       &      & 51.0  & 2.50  \\ \hline
           & $V^{ls}$ (MeV)   & $\mu^{\ell s}$ (fm$^{-2}$)   & $V_{\ell}^{\ell s}$ (MeV)   & $\mu_{\ell}^{\ell s}$ (fm$^{-2}$)    \\ 
Spin orbit & -16.8 & 0.52 & -22.0 & 0.396 \\ 
           &       &      & 20.0  & 2.200 \\ \hline  \hline
\end{tabular}
\label{KKNN}
\end{table}

\subsection{Variable Phase Approach}  
The Schr$\ddot{o}$dinger wave equation for a spinless particle with energy E and orbital angular momentum $\ell$ undergoing scattering is given by
\begin{equation}
\frac{\hbar^2}{2\mu} \bigg[\frac{d^2}{dr^2}+\big(k^2-\ell(\ell+1)/r^2\big)\bigg]u_{\ell}(k,r)=V(r)u_{\ell}(k,r)
\label{Scheq}
\end{equation}
Where $k=\sqrt{E/(\hbar^2/2\mu)}$.

Second order differential equation  Eq.\ref{Scheq} has been transformed to the first order non-homogeneous differential equation of Riccati type \cite{12,13} given by
\begin{equation}
\delta_{\ell}'(k,r)=-\frac{V(r)}{{k\left(\hbar^{2} / 2 \mu\right)}}\bigg[\cos(\delta_\ell(k,r))\hat{j}_{\ell}(kr)-\sin(\delta_\ell(k,r))\hat{\eta}_{\ell}(kr)\bigg]^2
\label{PFMeqn}
\end{equation}
 
Prime denotes differentiation of phase shift with respect to distance and the Riccati Hankel function of first kind is related to $\hat{j_{\ell}}(kr)$ and $\hat{\eta_{\ell}}(kr)$ by $\hat{h}_{\ell}(r)=-\hat{\eta}_{\ell}(r)+\textit{i}~ \hat{j}_{\ell}(r)$ .
Eq.\ref{PFMeqn} is numerically solved using Runge-Kutta 5$^{th}$ order (RK-5) method \cite{14} with initial condition $\delta_{\ell}(0) = 0$. For $\ell = 0$, the Riccati-Bessel and Riccati-Neumann functions $\hat{j}_0$ and $\hat{\eta}_0$ get simplified as $\sin(kr)$ and $-\cos(kr)$, so Eq.\ref{PFMeqn}, for $\ell = 0$ becomes 
\begin{equation}
\delta'_0(k,r)=-\frac{V(r)}{{k\left(\hbar^{2} / 2 \mu\right)}}\sin^2[kr+\delta_0(k,r)]
\end{equation}
In above equations $k=\sqrt{E/(\hbar^2/2\mu)}$ and $\hbar^2/2\mu$ = 25.9425 MeV.fm$^{2}$ for \textit{np} case.
In above equation the function $\delta_0(k,r)$ was termed ``Phase function'' by Morse and Allis \cite{15}.
Similarly by varying the Bessel functions for various $\ell$ values by using following recurrence relations \cite{16}
\begin{equation} 
\hat{j}_{{\ell}+1}(k r)=\frac{2 {\ell}+1}{k r} \hat{j}_{\ell}(k r)-\hat{j}_{{\ell}-1}(k r)\\ 
\end{equation}\begin{equation}
    \hat{\eta}_{{\ell}+1}(k r)=\frac{2 {\ell}+1}{k r} \hat{\eta}_{\ell}(k r)-\hat{\eta}_{{\ell}-1}(k r)
\end{equation}\newline  we obtain PFM equation for P-wave having following form

 \begin{equation}
    \delta_{1}^{\prime}(k, r)=\frac{-V(r)}{k\left(\hbar^{2} / 2 \mu\right)}\left[\frac{\sin \left(\delta_{1}+k r\right)-k r \cdot \cos \left(\delta_{1}+k r\right)}{k r}\right]^{2}
\end{equation}

The equation for  amplitude function\cite{17} with initial condition is obtained in the form

\begin{equation}
\begin{aligned}
    A_{\ell}^{\prime}(r) = &-\frac{A_{\ell} V(r)}{k} \left[\cos (\delta_\ell(k,r)) \hat{j}_{\ell}(kr)-\sin (\delta_\ell(k,r)) \hat{\eta}_{\ell}(kr)\right] \\
    &\times\left[\sin (\delta_\ell(k,r))( \hat{j}_{\ell}(kr)+\cos (\delta_\ell(k,r)) \hat{\eta}_{\ell}(kr)\right]
\end{aligned}
\end{equation}
\newline
also the equation to obtained wavefunction\cite{17} is 
\begin{equation}
    u_{\ell}(r)=A_{\ell}(r)\left[\cos (\delta_\ell(k,r)) \hat{j}_{\ell}(k r)-\sin (\delta_\ell(k,r)) \hat{\eta}_{\ell}(k r)\right]
\end{equation}

    \begin{figure}[ht!]
    \centering
    \includegraphics[scale=0.56]{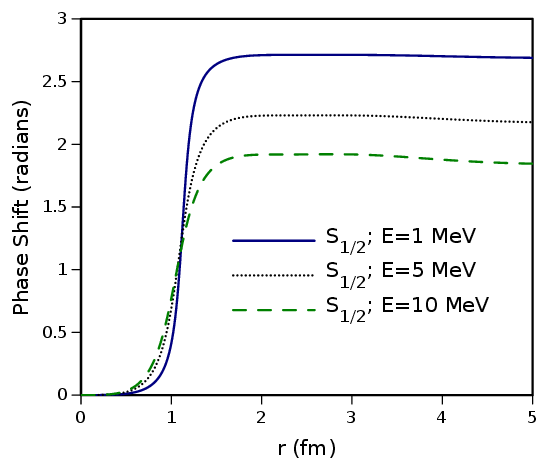}
    \includegraphics[scale=0.56]{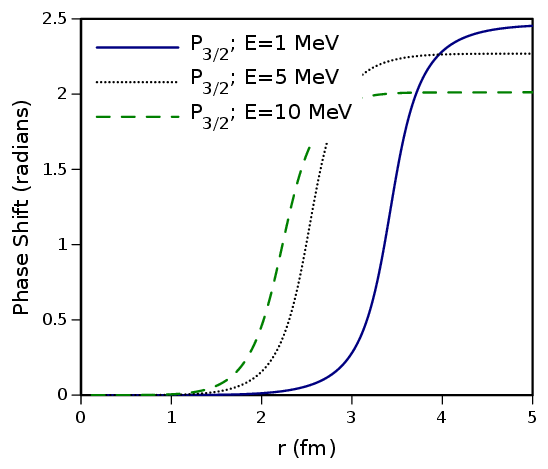}
    \includegraphics[scale=0.56]{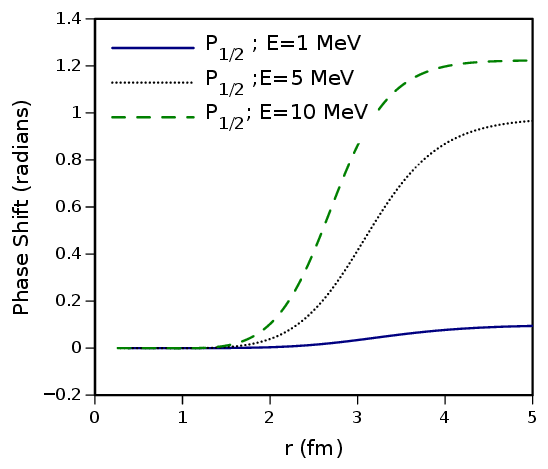}
    \includegraphics[scale=0.56]{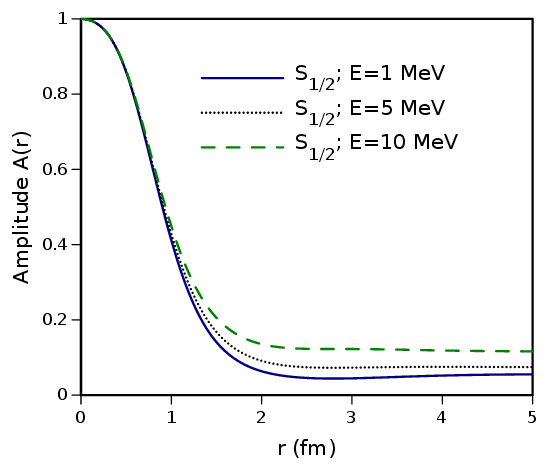}
    \includegraphics[scale=0.56]{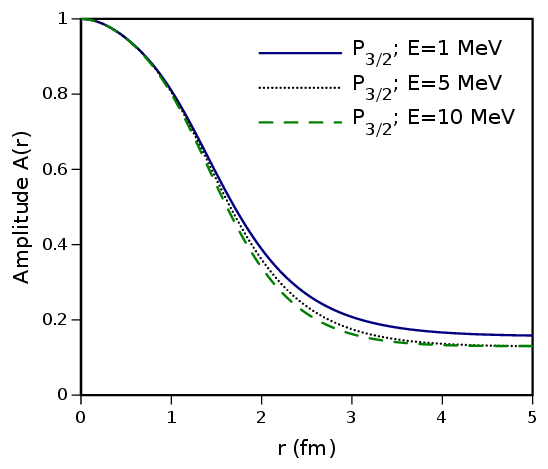}
    \includegraphics[scale=0.56]{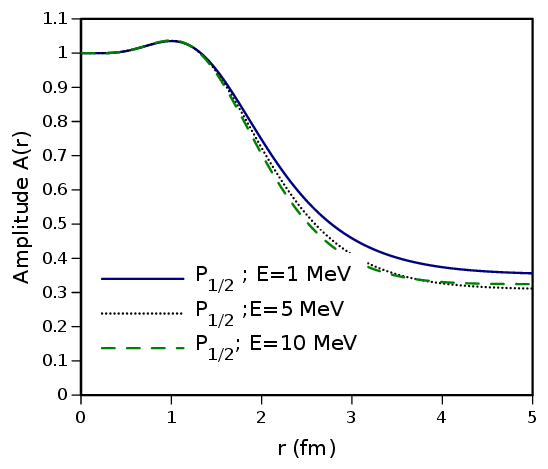}
    \includegraphics[scale=0.56]{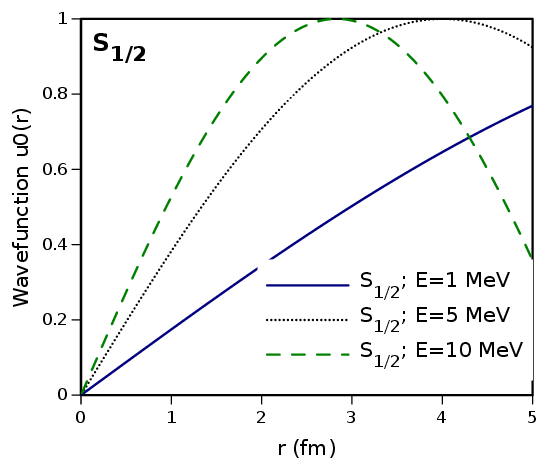}
    \includegraphics[scale=0.56]{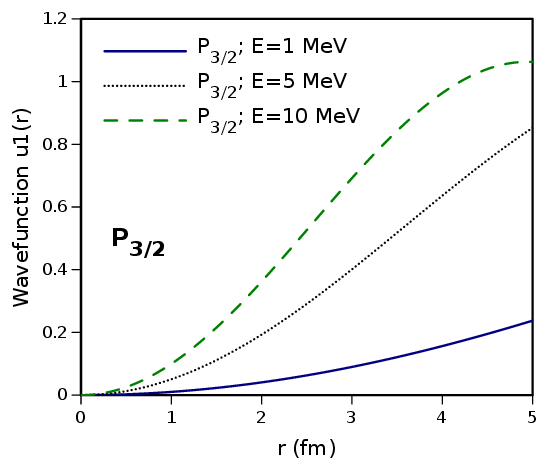}
    \includegraphics[scale=0.56]{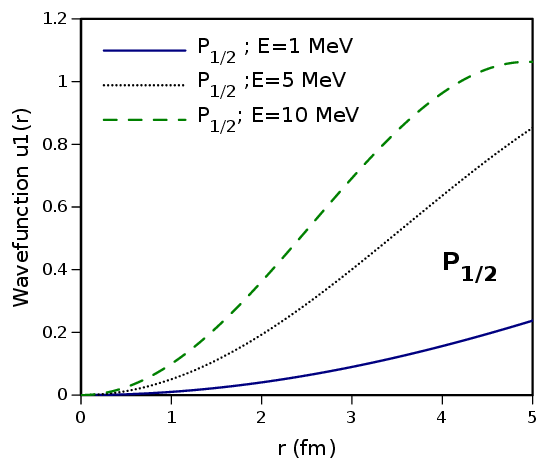}
   \caption{
Radial dependence of scattering quantities for neutron--alpha elastic
scattering at laboratory energies $E_{\text{lab}} = 1$, 5, and 10 MeV.
Top panels: distance-dependent phase shifts $\delta_\ell(r)$ showing the
gradual buildup of the scattering phase within the interaction region and
its saturation at large distances.
Middle panels: corresponding amplitude functions $A_\ell(r)$, which remain
finite and smooth over the entire radial domain.
Bottom panels: reconstructed radial wave functions $u_\ell(r)$ exhibiting
the expected nodal structure and asymptotic scattering behavior for each
partial wave.
}
    \label{fig:my_label}
    \end{figure}

    \section{Results and Discussion}
    Figure~1 shows the energy dependence of the scattering phase shifts for the
$S_{1/2}$, $P_{1/2}$, and $P_{3/2}$ partial waves along with the corresponding
effective interaction potentials. The sign and magnitude of the phase shifts
reflect the attractive or repulsive nature of the interaction in each channel.
The comparatively larger phase shifts observed in the $S_{1/2}$ channel
indicate the dominance of the central attractive potential at low energies.
In contrast, the reduced magnitude of the phase shifts in the $P$-wave channels
is a consequence of the centrifugal barrier, while the splitting between the
$P_{1/2}$ and $P_{3/2}$ states arises from the spin-orbit interaction.
The effective potential plots confirm the short-range character of the nuclear
interaction and demonstrate how the centrifugal term modifies the interaction
for higher partial waves.

The top panel of Fig.~2 displays the radial evolution of the phase shift $\delta_\ell(r)$ for different laboratory energies. The phase shift starts from zero at the origin, satisfying the boundary condition
$\delta_\ell(0)=0$, and increases within the region where the nuclear potential
is significant. Beyond the interaction range, $\delta_\ell(r)$ approaches a constant value,
corresponding to the physical scattering phase shift. 

The middle panel of Fig.~2 presents the amplitude functions $A_\ell(r)$ for
different energies and partial waves.
The amplitude functions remain finite and smooth over the entire radial domain,
demonstrating the numerical stability of the Runge-Kutta integration scheme.
Variations in $A_\ell(r)$ within the interaction region reflect the strength and
structure of the nuclear potential, while the stabilization of the amplitude at
larger distances indicates the transition to free-particle behavior outside the
range of the interaction. The bottom panel of Fig.~2 shows the radial wave functions $u_\ell(r)$ constructed
from the corresponding phase and amplitude functions.
\section{Conclusion} 
In this work, neutron-alpha elastic scattering has been investigated using the
Variable Phase Approach with the microscopic KKNN potential. Distance-dependent
phase shifts, amplitude functions, and radial wave functions were obtained for
the $S_{1/2}$, $P_{3/2}$, and $P_{1/2}$ partial waves over a range of laboratory
energies. The variable phase equations were solved numerically using a
fifth-order Runge-Kutta method, yielding stable and physically consistent
solutions.

The radial evolution of the phase shifts provides a transparent physical picture
of how the nuclear interaction generates the scattering phase within the finite
interaction region. The results clearly demonstrate the roles of the central
interaction, centrifugal barrier, and spin--orbit coupling in shaping the
scattering behavior of different partial waves. The obtained amplitude functions
remain well behaved, and the reconstructed wave functions exhibit the expected
nodal structure and asymptotic behavior, confirming the internal consistency of
the approach.

Overall, the present study shows that the Variable Phase Approach offers not only
computational efficiency but also enhanced physical insight into nucleon--nucleus
scattering phenomena. The methodology is well suited for inverse scattering
applications and can be readily extended to other systems such as proton-alpha
and neutron-nucleus scattering.

\section*{Appendix} 
\subsection*{Amplitude Function Equations}
\vspace{.1cm}
for $\ell=0$ \& $ \ell=1$ the amplitude equations are:
\vspace{.1cm}
\begin{align}\nonumber
A_{0}^{\prime} &= -\frac{A_{0} V(r)}{k\left(\frac{\hbar^{2}}{2\mu}\right)} \left[\cos \delta_{0} \cdot \sin(kr) - \sin \delta_{0} \cdot (-\cos(kr))\right] \\
&\quad \times \left[\sin \delta_{0} \cdot \sin(kr) + \cos \delta_{0} \cdot (-\cos(kr))\right] \\
\\
A_{1}^{\prime} &= -\frac{A_{1} V(r)}{k\left(\frac{\hbar^{2}}{2\mu}\right)} \left[ \cos \delta_{1} \left( \frac{\sin (kr)}{(kr)}-\cos (kr) \right) - \sin \delta_{1} \left( -\frac{\cos (kr)}{(kr)}-\sin (kr) \right) \right] \\
&\quad \times \left[ \sin \delta_{1} \left( \frac{\sin (kr)}{(kr)}-\cos (kr) \right) + \cos \delta_{1} \left( -\frac{\cos (kr)}{(kr)}-\sin (kr) \right) \right]
\end{align}
\subsection*{Wavefunction Equations}
\vspace{.1cm}
for $\ell=0$ \& $\ell=1$ the wavefunction equations are: 
\vspace{.1cm}
\begin{flalign}
    &u_{0}(r) = A_{0}(r) \left[\cos \delta_{0}(r) \cdot \sin(kr) - \sin \delta_{0}(r) \cdot \cos(kr)\right] & \\
    &u_{1}(r) = A_{1}(r) \left[\cos \delta_{1}(r) \left(\frac{\sin (kr)}{(kr)}-\cos (kr)\right) - \sin \delta_{1}(r) \left(-\frac{\cos (kr)}{(kr)}-\sin (kr)\right)\right] & \\
\end{flalign}

\end{document}